%% file: nplusSPT_V3.tex
\documentclass[fleqn,usenatbib]{mnras}

\usepackage{graphicx}
\usepackage{subfig} 
\usepackage{times}
\usepackage{hyperref}
\usepackage[T1]{fontenc}
\usepackage{aecompl}
\usepackage{soul}

\newcommand{\CII}{\mbox{[C~{\sc ii}]}}
\newcommand{\NII}{\mbox{[N~{\sc ii}]}}
\newcommand{\OI}{\mbox{[O~{\sc i}]}}
\newcommand{\HII}{\mbox{H~{\sc ii}}}
\newcommand{\median}{11.0}
\newcommand{\survmedian}{9.7}

\newcommand{\interquartile}{5.0 to 24.7}
\newcommand{\survinterquartile}{5 to 25}

\newcommand{\FIR}{\sc fir}
\newcommand{\ratioCN}{L$_{\textrm{\small \CII }}$/L$_{\textrm{\small \NII }}$}

\begin{document}

\title[The \CII\ /\NII\ distribution in high-z SPT SMGs]
{The \CII\ / \NII\ ratio in $3<z<6$ sub-millimetre galaxies   from the South Pole Telescope survey}

\author[D. J. M. Cunningham et al.]
{D. J. M. Cunningham$^{1,2}$, S. C. Chapman$^{2,3,4}$, M. Aravena$^5$, C. De Breuck$^6$, M. B\'{e}thermin$^7$,
\newauthor Chian-Chou Chen$^6$, Chenxing Dong$^8$, A. H. Gonzalez$^8$, T. R. Greve$^{9,18}$, K. C. Litke$^{10}$, 
\newauthor J. Ma$^8$, M. Malkan$^{11}$, D. P. Marrone$^{10}$, T. Miller$^{2,12}$, K. A. Phadke$^{13}$, C. Reuter$^{13}$, 
\newauthor K. Rotermund$^2$, J. S. Spilker$^{14}$, A. A. Stark$^{15}$, M. Strandet$^{16,17}$, J. D. Vieira$^{13}$, 
\newauthor A. Wei\ss{}$^{16}$.
\\
$^1$Department of Astronomy and Physics, Saint Mary's University, Halifax, NS, B3H 3C3, Canada\\
$^2$Department of Physics and Atmospheric Science, Dalhousie University, Halifax, NS, B3H 4R2, Canada\\
$^3$National Research Council, Herzberg Astronomy and Astrophysics, Victoria, British Columbia, Canada\\
$^4$Department of Physics and Astronomy, University of British Columbia, Vancouver, BC, V6T 1Z1, Canada\\
$^5$N\'ucleo de Astronom\'ia, Facultad de Ingenier\'ia, Universidad Diego Portales, Av. Ej\'ercito 441, Santiago, Chile\\
$^6$European Southern Observatory, Karl Schwarzschild Straße 2, 85748 Garching, Germany\\
$^7$Aix Marseille University, CNRS, LAM, Laboratoire d'Astrophysique de Marseille, Marseille, France\\
$^8$Department of Astronomy, University of Florida, Bryant Space Sciences Center, Gainesville, FL 32611 USA\\
$^9$Department of Physics and Astronomy, University College London, Gower Street, London WC1E 6BT, UK\\
$^{10}$Steward Observatory, University of Arizona, 933 North Cherry Avenue, Tucson, AZ 85721, USA\\
$^{11}$Department of Physics and Astronomy, University of California, Los Angeles, CA 90095-1547, USA\\
$^{12}$Department of Astronomy, Yale University, New Haven, CT, USA\\
$^{13}$Department of Astronomy, University of Illinois, 1002 West Green St., Urbana, IL 61801\\
$^{14}$Department of Astronomy, University of Texas at Austin, 2515 Speedway, Stop C1400, Austin, TX 78712, USA\\
$^{15}$Harvard-Smithsonian Center for Astrophysics, 60 Garden Street, Cambridge, Massachusetts 02138, USA\\
$^{16}$Max-Planck-Institut für Radioastronomie, Auf dem Hügel 69, D-53121 Bonn, Germany\\
$^{17}$International Max Planck Research School (IMPRS) for Astronomy and Astrophysics, Universities of Bonn and Cologne, Bonn, Germany\\
$^{18}$Cosmic Dawn Center (DAWN), DTU-Space, Technical University of Denmark, Elektrovej 327, DK-2800 Kgs.\\
Lyngby; Niels Bohr Institute, University of Copenhagen, Juliane Maries Vej 30, DK-2100 Copenhagen $\O$\\
}
\date{draft 1.0}

\maketitle
\begin{abstract}
\noindent
We present Atacama Compact Array and Atacama Pathfinder Experiment observations of the \NII\ 205 $\mu$m fine-structure line in 40 sub-millimetre galaxies lying at redshifts $z=3$ to $6$, drawn from the 2500 deg$^2$ South Pole Telescope survey.
This represents the largest uniformly selected sample of high-redshift \NII\ 205 $\mu$m measurements to date. 
29 sources also have \CII\ 158 $\mu$m line observations allowing a characterization of the distribution of the \CII\ to \NII\ luminosity ratio for the first time at high-redshift.
The sample exhibits a median \ratioCN\ $\approx$ \median\ and interquartile range of \interquartile.
These ratios are similar to those observed in local (U)LIRGs, possibly indicating similarities in their interstellar medium.
At the extremes, we find individual sub-millimetre galaxies with \ratioCN\ low enough to suggest a smaller contribution from neutral gas than ionized gas to the \CII\ flux and high enough to suggest strongly photon or X-ray region dominated flux.
These results highlight a large range in this line luminosity ratio for sub-millimetre galaxies, which may be caused by variations in gas density, the relative abundances of carbon and nitrogen, ionization parameter, metallicity, and a variation in the fractional abundance of ionized and neutral interstellar medium.
\end{abstract}


\footnotetext{E-mail: dcunningham@dal.ca}

\section{Introduction}

Observing far-infrared (FIR) luminous galaxies at high-redshift is a crucial step in understanding the evolution of galaxies as they highlight periods of intense star formation which may represent pivotal growth periods in a galaxy's evolution (e.g., \protect\citealt{casey14}).
At z $\sim$ 2, sub-millimetre galaxies (SMGs) may have accounted for 50\% of star formation in the universe \protect\citep{wardlow11}.
At high-redshift, the FIR emission from these galaxies peaks at sub-millimetre wavelengths in the observer's frame, allowing effective selection at sub-mm or longer wavelengths.
The peak redshift at which they are predominantly found depends on the wavelength of the survey: at $\sim$ 850$\mu$m, the peak is close to  $z\sim2.5$ \protect\citep{chapman03, chapman05}, while at the 1 to 2 mm regime of the SPT survey, the median redshift increases to $\langle z\rangle \sim 4$ (\protect\citealt{weiss13}, \protect\citealt{strandet16}).
Regardless of wavelength of selection, they display rapid star formation sometimes exceeding $10^{3}$ $\mathrm{M}_{\odot}/\mathrm{yr}$ \protect\citep{swinbank14}, and have stellar masses on the order of $10^{11}$ $\mathrm{M}_{\odot}$ \protect\citep{hainline11, michalowski12, ma15}.
Their rapid evolution early in cosmic time continues to push current simulations to match their detailed properties (e.g. \protect\citealt{shimizu12, hayward13, narayanan15, cowley17}).

\begin{figure}
	\centering
	\subfloat{{\includegraphics[width=0.45\textwidth]{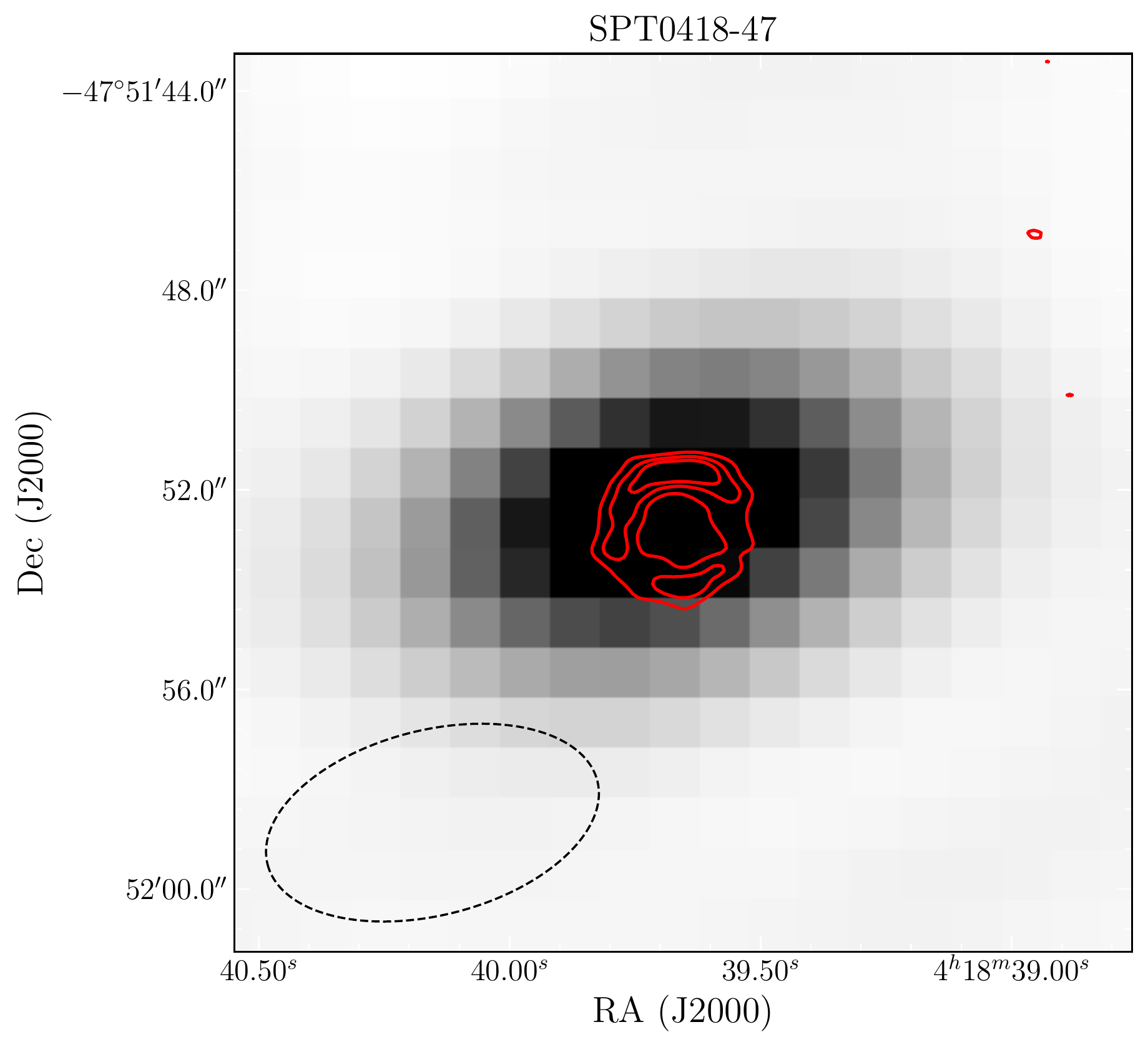}}}\\
	\vskip-0.1cm \subfloat{{\includegraphics[width=0.45\textwidth]{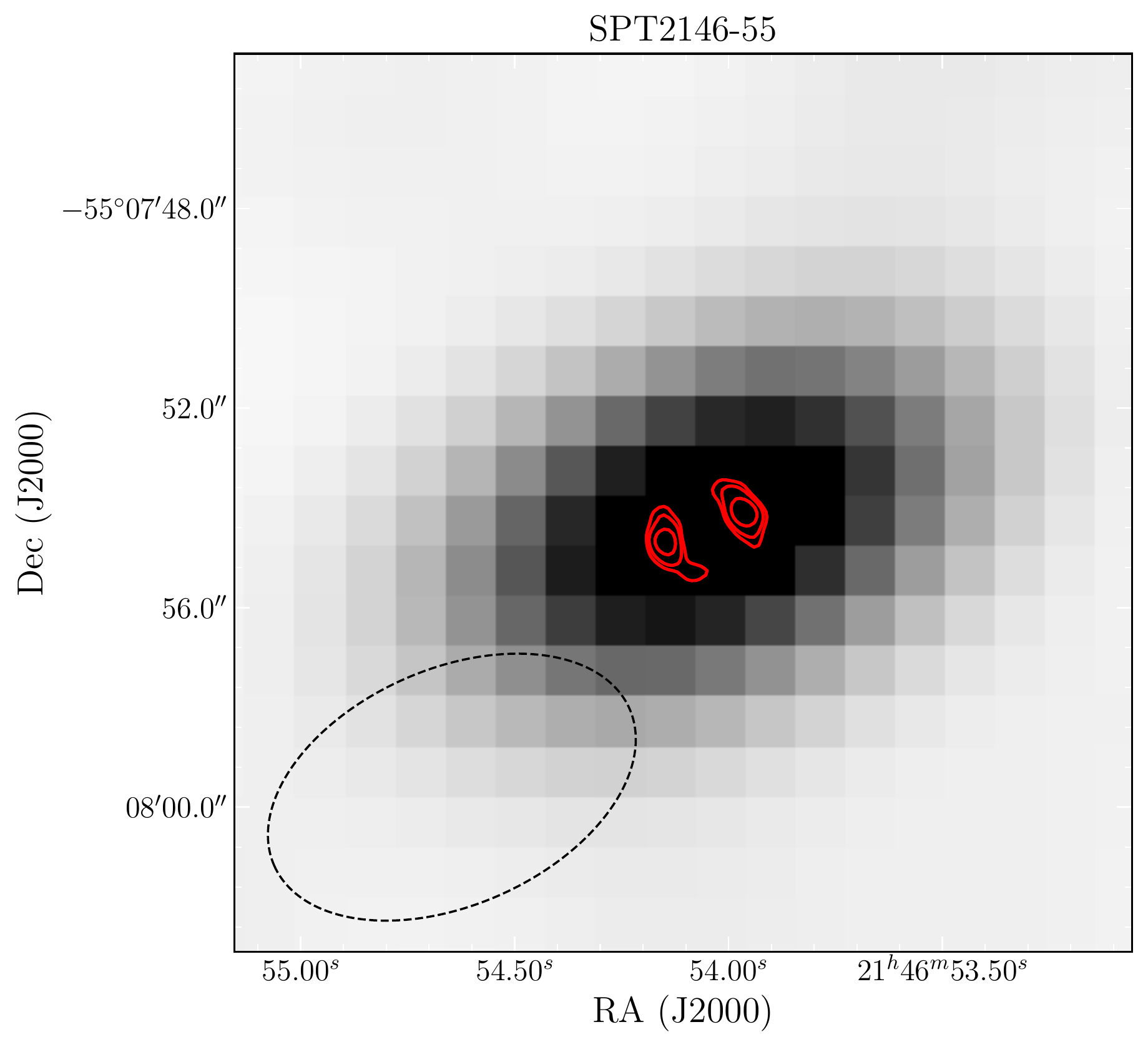}}}
	\caption{Two examples of our ALMA ACA \NII\ line observations, one from each of band 6 and 7, showing the \NII\ detection as a channel map optimized over the FWHM of the line in grey-scale. The red contours represent continuum from higher resolution band-7 ALMA imaging, and demonstrate that the ACA beam does not resolve even the largest lensed SMGs in our sample. The beam size is represented in the bottom left corner by the black dashed line ellipse. \textit{Top panel:} \NII\ observations of SPT0418--47 in band-7. \textit{Bottom panel:} \NII\ observations of SPT2146--55 in band-6.}
	\label{fig:fig2}
\end{figure}

In high-redshift dusty galaxies more traditional optical and ultraviolet line diagnostics are not possible due to high dust attenuation.
Fine-structure transition lines such as \CII\ 158 $\mu$m (${}^{2}\mathrm{P}_{3/2} \rightarrow {}^{2}\mathrm{P}_{1/2}$) and \NII\ 205 $\mu$m (${}^{3}\mathrm{P}_{1} \rightarrow {}^{3}\mathrm{P}_{0}$) (hereafter \CII\ and \NII) offer important insight into the properties of the interstellar medium (ISM) and importantly are not significantly affected by dust attenuation.
The \CII\/-to-\NII\ ratio can probe physical parameters of the ISM.
Assuming a pressure-equilibrium gas cloud with a range of gas densities and ionization parameters at its illuminated surface, \protect\cite{nagao12} used \textsc{cloudy} modelling \protect\citep{ferland98} to show that the \CII\/-to-\NII\ flux ratio decreases monotonically with gas metallicity.
However, because of dependencies of this ratio on the unknown density and ionization parameters, additional lines such as \NII\ 122 $\mu$m and \OI\ 145 $\mu$m are needed to break the degeneracy with these parameters \protect\citep{nagao12}.
Since \CII\ emission originates from both ionized and neutral gas, while \NII\ is primarily emitted from ionized gas, the \CII\/-to-\NII\ ratio can probe the abundance of ionized and neutral gas regions in a galaxy's ISM \protect\citep{decarli14}.

Several previous studies have used the \CII\/-to-\NII\ line ratio to investigate the ISM properties of luminous, high-redshift galaxies including SMGs (e.g. \protect\citealt{nagao12}, \protect\citealt{decarli14}, \protect\citealt{bethermin16}, \protect\citealt{pavesi16}, \protect\citealt{umehata17}, \protect\citealt{pavesi18}, \protect\citealt{tadaki19}).
They measure a large range in the line ratio, indicating these galaxies have diverse ISM conditions. 
At low-redshift, 
\protect\cite{herreracamus16} used \NII\ 122 $\mu$m and 205 $\mu$m emission lines to constrain gas density and SFR, while \protect\cite{cormier15} found (using the \NII\ 122 $\mu$m line) that the ionized medium contributes little to the \CII\ emission in their dwarf galaxy sample.

This paper presents measurements of \NII\ in 40 gravitationally lensed SMGs between $3<z<6$ from the SPT survey.
This is the first uniformly selected large sample of SMGs with \NII\ detections at high-redshift. 
When combined with 29 additional \CII\ observations, these measurements allow us to make the first characterization of the high-redshift \ratioCN\ distribution in SMGs using a uniformly selected sample.
Its wide redshift range makes it a unique sample to study the possible evolution in the ISM of high-redshift SMGs in comparison to local Luminous Infra-red Galaxies (LIRGs). 
Since the SPT galaxies are gravitationally lensed, even the relatively faint \NII\ emitting sources can be detected quickly with the Morita Atacama Compact Array (ACA) of the Atacama Large Millimeter/submillimeter Array (ALMA). 
This allows for a complete characterization of the \ratioCN\ ratio in the ultra-luminous galaxy population (L$_{\textrm{\FIR }}>10^{12}$ L$_\odot$) at high-redshift.
We assume a Hubble constant $H_0=70$\ km\,s$^{-1}$\,Mpc$^{-1}$ and density parameters $\Omega_{\Lambda}=0.7$ and $\Omega_{\rm m}=0.3$ throughout.

\begin{figure*}
	\centering
	\includegraphics[width=\textwidth]{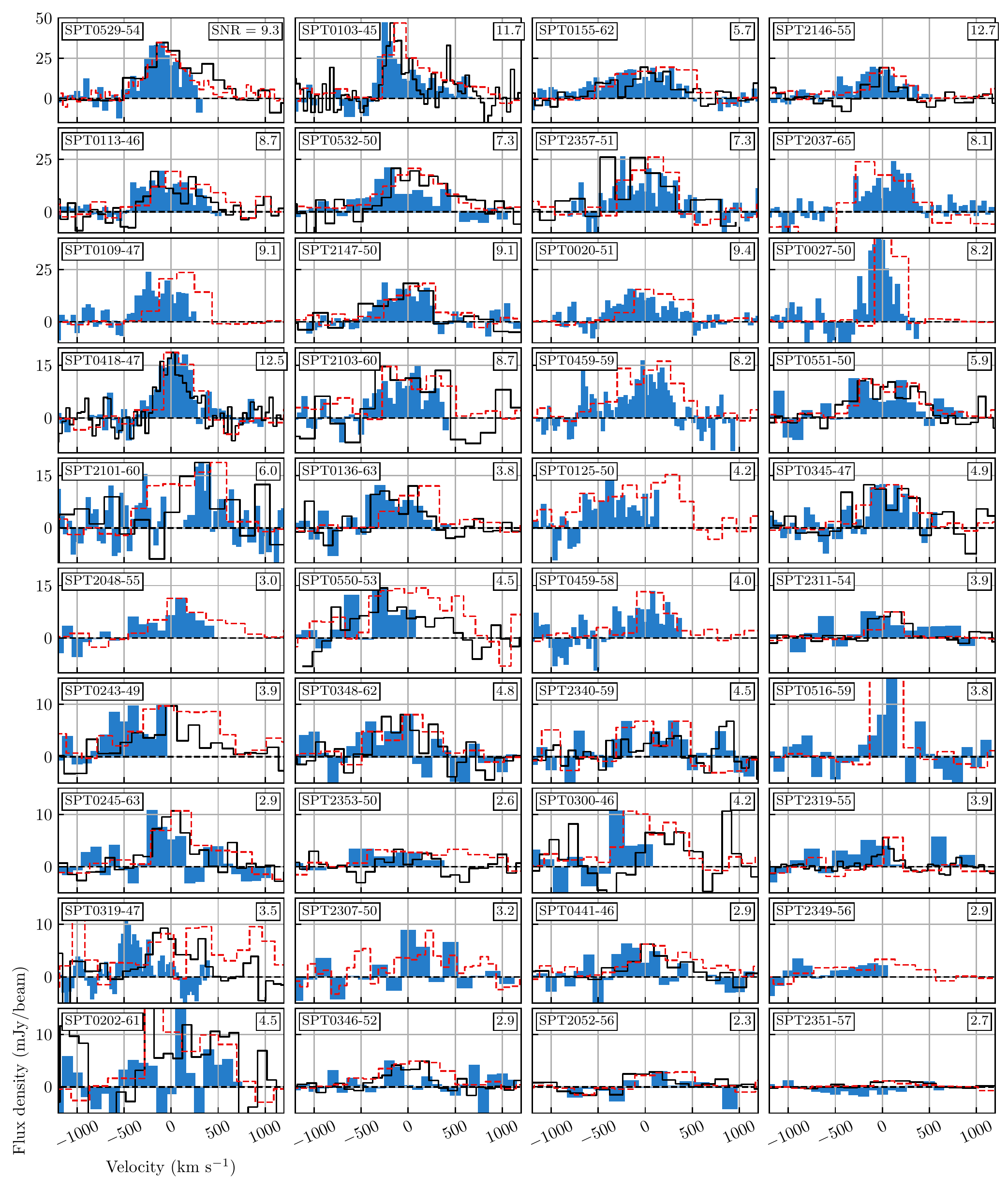}
	\caption{Observed \NII\ 205 $\mu$m (blue background), \CII\ 158 $\mu$m (black line), and CO (dashed red line) line profiles for the 40 ALMA observed sources, excluding SPT2132--58 which was presented in \protect\cite{bethermin16}.
Sources are ordered by observed \NII\ 205 $\mu$m flux, given in Table~\ref{table:tab1}.
The \NII\ spectra are single pixel extractions.
The signal-to-noise ratio from the optimal integrated line detection is given in the upper-right corner of each panel.
The CO line is either CO(4--3) or CO(5--4) depending on source redshift (see \protect\citealt{strandet16}).
Both CO and \CII\ lines are normalized to the peak \NII\ flux for each source (see Table~\ref{table:tab1} for \CII\ flux).
The zero velocity is determined according to redshifts presented in \protect\cite{strandet16}.}
	\label{fig:fig1}
\end{figure*}

\begin{figure*}
	\centering
	\includegraphics[width=\textwidth]{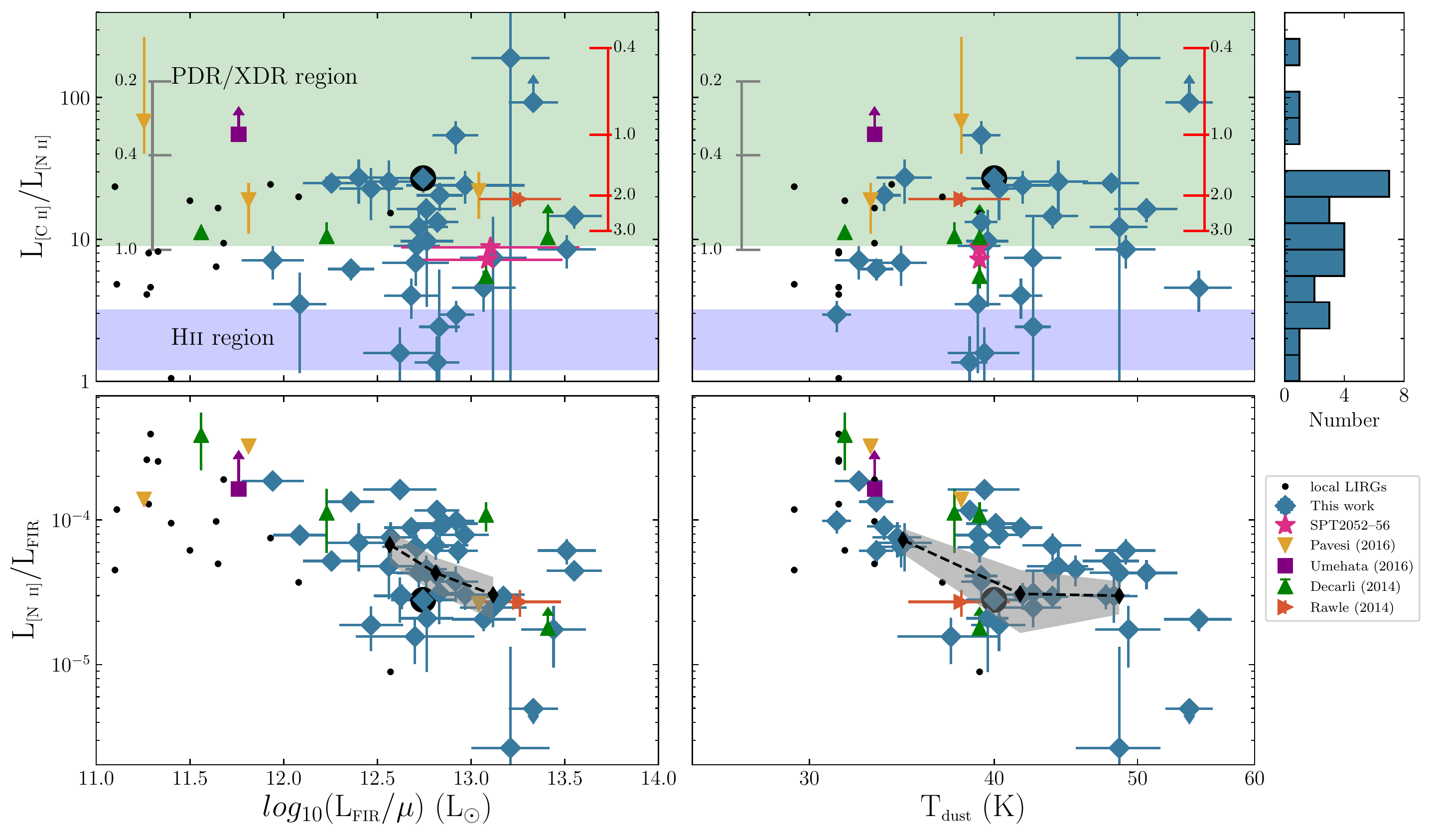}
	\caption{Comparison of key properties of our lensed sample of \NII\ and \CII\ observations (shown in blue diamonds) with literature measurements. 
\textit{Top panels:} Observed \ratioCN\ plotted against de-magnified L$_{\textrm{\small FIR}}$ and dust temperature.
Expected \ratioCN\ regimes for \HII\ and PDR/XDR zones based on \protect\cite{decarli14} are shown.
Modelled metallicity grids for ISM densities of $\log(n)=1.5$ (grey) and $3.0$ (red), each with ionization parameter $\log($U$_{\textrm{\small \HII }})=-3.5$ are shown as floating axes (\protect\citealt{nagao12}) with metallicity units of $Z/Z_{\odot}$.
\textit{Bottom panels:} Observed L$_{\textrm{\small FIR}}$ normalized \NII\ luminosities.
The black diamonds and the grey shaded regions in the bottom panels represent the binned medians and errors of de-magnified L$_{\textrm{\small FIR}}$ normalized \NII\ luminosities and dust temperature.
For literature sources, the upwards green triangles represent sources from \protect\cite{decarli14}: two Lyman alpha emitters, a SMG, and a QSO in order of increasing $L_{\textrm{\FIR }}$ at $z=4.7$.
The downwards yellow triangles represent \protect\cite{pavesi16} sources, which are LBG-1 (typical lyman-break galaxy), HZ10 (FIR-luminous LBG), and AzTEC-3 (a massive SMG) in order of increasing $L_{\textrm{\FIR }}$ at $z=5.3$ to $5.65$.
The purple square represents LAB1-ALMA3, a galaxy embedded in a Lyman alpha blob at $z=3.1$ \protect\citep{umehata17}.
A SMG (rightward orange triangle) at $z=5.243$ from \protect\cite{rawle14}, and two components of SPT2052--56, an unlensed proto-cluster of SMGs at $z=4.3$ are also shown (pink stars -- Pass et al.\ in prep).
SPT2132--58 \protect\citep{bethermin16} has a black circular background.
Local LIRGs assembled from \protect\cite{lu17}, \protect\cite{diazsantos17}, and \protect\cite{zhao16} are shown as small black dots.}

	\label{fig:fig3}
\end{figure*}

\begin{figure}
	\centering
	\subfloat{{\includegraphics[width=0.45\textwidth]{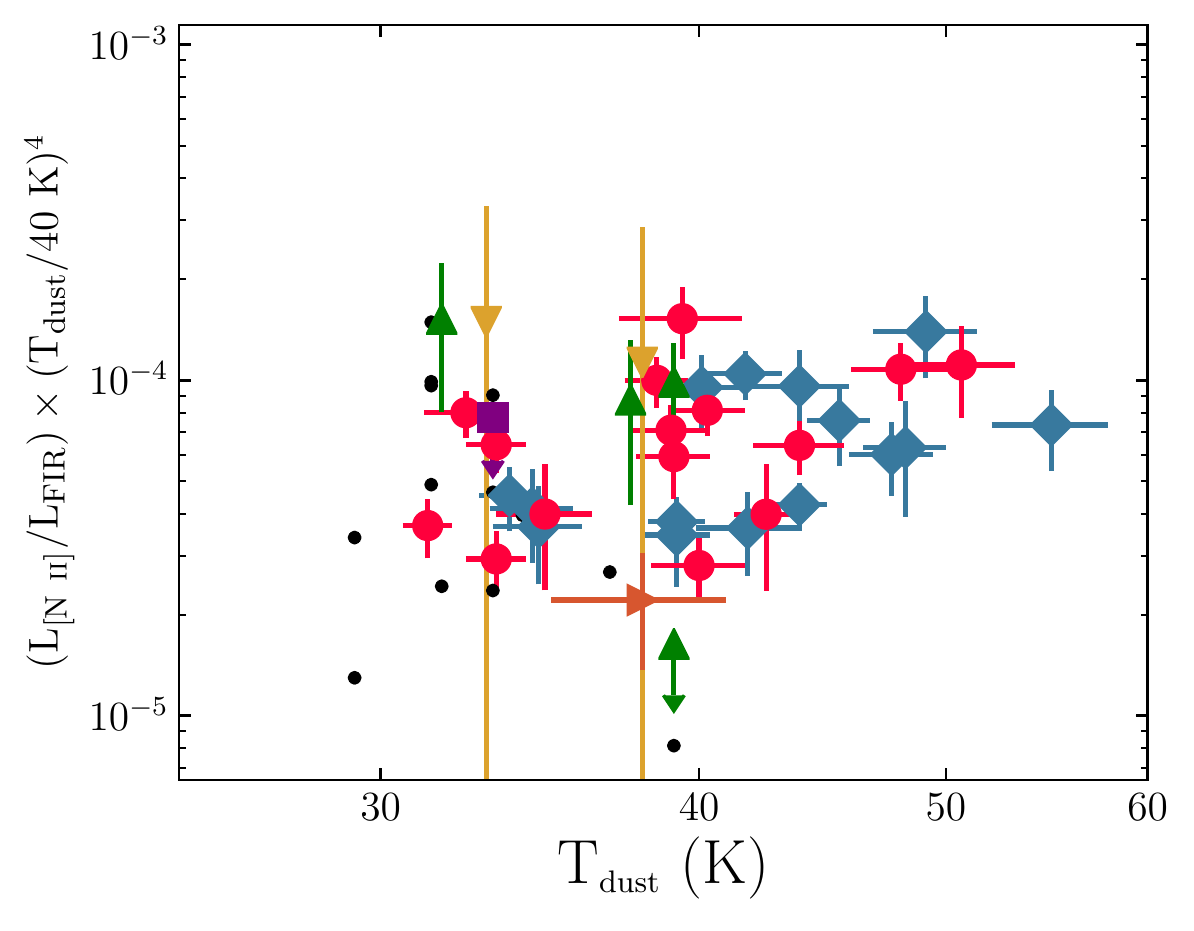}}}\\
	\caption{The distribution of $($L$_{\textrm{\small \NII }}$/L$_{\textrm{\small FIR}})\times ($T$_{\textrm{\small dust}}/40$ K$)^4$ versus T$_{\textrm{\small dust}}$. We plot SPT SMGs in blue diamonds and red circles. Blue diamonds represent sources with \ratioCN\ $>10$, red circles represent \ratioCN\ $<10$. This allows us to investigate whether sources with more significant \NII\ emission exhibit a relationship between L$_{\textrm{\small \NII }}/$L$_{\textrm{\small FIR}}$ and dust temperature. The other souces utilize the same legend as Figure~\ref{fig:fig3}.}
	\label{fig:fig4}
\end{figure}

\section{Sample selection, observations and data reduction}

This sample of 40 gravitationally lensed SMGs are selected from the South Pole Telescope Sunyaev-Zeldovich (SPT-SZ) survey (\protect\citealt{vieira10}, \protect\citealt{mocanu13}) covering 2500 deg$^{2}$ at 3, 2, and 1.4 mm wavelengths.
We selected a subset of sources with $3<z<6$ in order to enable observation of the \NII\ 205 $\mu$m line within bands 6 and 7 of ACA, including many sources which had existing \CII\ observations (e.g. \protect\citealt{gullberg15}).
Each source has a secure spectroscopic redshift (Table~\ref{table:tab1}) determined primarily using CO transitions and other fine-structure lines (see \protect\citealt{strandet16} for details).

The \NII\ line observations for all sources  were observed with ACA in Cycle 4 (PI: Chapman, 2016.1.00133.T), except for SPT2132--58 which was observed with the ALMA 12 m array (\citealt{bethermin16}).
ACA is ideal for our measurements of the total \NII\ line flux from lensed sources with large (up to $\sim1.5''$) Einstein radii because its FWHM beam size is 3 to 4$\arcsec$ at these frequencies, with the short observations typically yielding elongated beams due to restricted UV coverage.
The ACA sensitivity is still sufficient to detect this relatively faint \NII\ line (compared to the much brighter \CII) in the gravitationally lensed SMGs. 
The \CII\ observations used here were taken with the single dish Atacama Pathfinder Experiment (APEX) and are described in detail in \protect\cite{gullberg15}.

The \NII\ data were reduced using the Common Astronomy Software Applications package (\textsc{casa}) version 4.7 \protect\citep{mcmullin07}. \textsc{casa}'s \textsc{clean} function was used to generate continuum images and line cubes. The \textsc{clean} depth varied depending on the source, but was between 2 to 5$\sigma$.
The typical pixel size was 1\arcsec\ with beam semi-major (minor) axes of approximately 5 to 7\arcsec\ (3 to 4)\arcsec.
Our observations of \NII\ and \CII\ lines are shown in Figure~\ref{fig:fig1}.
Our observations achieved RMS continuum noise of 0.5 to 0.9 mJy per beam.

Figure~\ref{fig:fig2} illustrates continuum images of two of the most extended lensed SMGs in our sample, one from each of band 6 and 7, with high-resolution ALMA band 7 continuum contours superposed.
This figure illustrates that our ACA observations are unresolved even for the largest sources.
To test this assumption, we extracted both continuum and line flux from elliptical aperture regions corresponding to 1$\times$ and 1.5$\times$ the beam size, in addition to a point source, single pixel extraction.
These aperture extractions did not alter (or increase) the line flux measurement, indicating complete flux contained within the extraction pixel.
\NII\ spectral lines and the corresponding continuum were extracted at the peak of emission for all sources, with a single pixel extraction, and are shown in Figure~\ref{fig:fig1} with integrated line fluxes and luminosities given in Table~\ref{table:tab1}.
The continuum emission is subtracted using a zeroth order polynomial matched to the flux baseline of off-line regions.

In Figure~\ref{fig:fig1}, we plot \NII\,, \CII\,, and either CO(5-4) or CO(4-3) from \protect\cite{strandet16}, depending on redshift. 
In order to most reliably determine the integrated fluxes in both \NII\ and \CII, we use the FWHM$_{\textrm{\small CO}}$ of the high signal-to-noise ratio (SNR) CO lines to determine the velocity range over which to sum our \NII\ and \CII\ detections.
We first fit the CO line profiles with a single Gaussian function and then use this to obtain our line flux measurements by summing \NII\ and \CII\ over a velocity range $2\times$FWHM$_{\textrm{\small CO}}$, covering the full-width of our \NII\ line profiles.
For brighter \NII\ sources we confirmed through a curve of growth analysis that this represents $>95$\% of the line flux, while for fainter \NII\ sources this avoids large variations in the line flux measurement from integrating noise fluctuations outside the line frequencies.
For sources with both \NII\ and \CII\ observations, we compared the FWHM determined from a Gaussian fit for each line.
Over the full sample, there is a good one-to-one agreement between these two line widths, although the relatively low SNRs of many of the lines (both \NII\ and \CII\/) result in significant scatter in the relation.

We fit the \NII\ line profiles and compare to the \CII\ profiles, both listed in Table~\ref{table:tab1}.
In 4 of the 40 cases, our spectral bandwidth only covers between 50\% to 70\% of the CO-defined line (SPT0125--50, SPT0300--46, SPT0243--49, and SPT0550--53).
This was a compromise taken in order to attempt to reduce the overall project calibration overhead, since the total project exceeded 50 hours.
In these cases, we only sum the \CII\ line to the end of the \NII\ line coverage.
The \NII\ and \CII\ fluxes, along with the observed \NII\ luminosities are presented in Table~\ref{table:tab1}.
In Figure~\ref{fig:fig3}, the demagnified \NII\ luminosities are plotted against demagnified L$_{\textrm{\small FIR}}$ and T$_{\textrm{\small dust}}$, using magnification values ($\mu$) from \protect\cite{spilker16}.

\section{Results}
\label{sec:results}

We robustly detect \NII\ lines with the ACA at $>4.5\sigma$ in 22 of our sample of 41 SMG observations (including SPT2132-58 from \protect\citealt{bethermin16}), with a further 13 sources detected at $\geq3\sigma$.
The $>4.5\sigma$ criteria results in a 5\% false positive chance in our data, which drops to less than 1\% chance when the measurement is constrained to within one-half beam of the CO-source coordinates.
These lines are detected at the redshifted velocity expected based on redshifts presented in \protect\cite{strandet16}.
Along with the $11.5\sigma$ \NII\ detection of SPT2132--58 (\protect\citealt{bethermin16}), this represents a total of 35 detections in the sample of 41.
In the 6 sources with undetected \NII, an upper limit was estimated as 3$\times$ the channel RMS in the central pixel at velocities away from the line.
For our lines with $\sigma \geq3$, we find magnified \NII\ line fluxes ranging between approximately $1$ to $14\textrm{~Jy km/s}$ and \CII\ flux between $5$ to $216\textrm{~Jy km/s}$.
These correspond to \NII\ line luminosities of $(\sim 5$ to $40)\times 10^{8}$ L$_{\odot}$, and detected \ratioCN\ line luminosity ratios over two orders of magnitude.
The sample has an interquartile range in line luminosity ratios \ratioCN\ of \interquartile\ and a median of \median\ (see Figure~\ref{fig:fig3}, rightmost panel for a histogram detailing this distribution).
These statistical values are calculated by treating the \ratioCN\ lower limits (\NII\ non-detections) as real measured values at the measured limit.

To properly account for the lower limits in \ratioCN\,, we perform survival analysis using the \textsc{lifelines} Python package \protect\citep{lifelines}.
Survival analysis is often used to determine the time until an event occurs.
In cases where an event is not precisely observed, the last observation made before the event occurs can still be used as a lower limit in calculating statistical quantities.
For our line ratios, we utilize the lower limits as the last observation before the ``event'' occurs, where the event is the true line ratio.
This analysis gives a slightly lower median value of \survmedian\ with a similar interquartile range of \survinterquartile\/.
Furthermore, reducing our sample to include only sources with good quality \NII\ and \CII\ detections did not significantly alter our measured medians or interquartile ranges.

In the majority of our sources ($17/30$, or $\sim$ 60 \%), the \ratioCN\ luminosity ratio (or lower limit) corresponds to model expectations from XDR/PDR or shock regions determined by \protect\cite{decarli14}.
\protect\cite{decarli14} use L$_{\textrm{\small \CII }}/$L$_{\textrm{\small \NII }}\sim 2$ for \HII\ regions and greater than L$_{\textrm{\small \CII }}/$L$_{\textrm{\small \NII }}\sim 9$ for PDR/XDR regions (see Figure~\ref{fig:fig3}).
Three SPT SMGs fall well within the range expected for \HII-dominated regions (with another 3 overlapping within error or as a lower limit), with the rest existing in an intermediate region, or more balanced regime in between the ionized gas dominated and PDR/shock dominated regions.
Galaxies with \CII/\NII\ ratios in the XDR/PDR or shock region regime are expected to have \NII\ emission originating predominantly from \HII\ regions with \CII\ emission originating from both \HII\ regions and the outer layers of PDRs \protect\citep{bethermin16}.

We assembled a sample of \ratioCN\ in local LIRGs from \protect\cite{diazsantos17}, \protect\cite{lu17}, and \protect\cite{zhao16}.
To investigate whether the SPT SMG sample and the local LIRGs arise from different underlying distributions of \ratioCN{}, we perform a two-sample KS test and T-test.
The KS test yields a \textit{p}-value of 0.44, while the T-test yields \textit{p} $=0.36$.
These test results prevent us from conclusively determining the line ratios come from different underlying distributions.
In our literature sample, when dust temperatures were not available they were estimated using the relationship shown in \protect\cite{symeonidis13}.

We investigate the relationship between L$_{\textrm{\small \NII }}/$L$_{\textrm{\small FIR}}$ versus L$_{\textrm{\small FIR}}$ (42.5 to 122.5 $\mu$m) by binning our sources according to L$_{\textrm{\small FIR}}$.
We divide our sources with L$_{\textrm{\small \NII }}$ measurements into three bins of roughly equal size, ranging from log$_{10}($L$_{\textrm{\small FIR}})=$[$<$12.71, 12.71--12.9,$>$12.9$]$.
From smallest to largest bins, we measure L$_{\textrm{\small \NII }}/$L$_{\textrm{\small FIR}}$ of $\sim 6.8\times 10^{-5}$, $4.3\times 10^{-5}$, and $3.0\times 10^{-5}$. These bins and their medians (and errors) are shown as black diamonds and grey shaded regions in the lower-left panel of Figure~\ref{fig:fig3}.

We also investigate the trend between dust temperature and L$_{\textrm{\small \NII }}$/L$_{\textrm{\small FIR}}$.
These results are presented in the bottom-right panel of Figure~\ref{fig:fig3}, where we plot L$_{\textrm{\small \NII }}$/L$_{\textrm{\small FIR}}$ against T$_{\textrm{\small dust}}$.
We also bin our sources into 3 bins of roughly equal number according to dust temperature: T$_{\textrm{\small dust}}=$[$<$39.3, 39.3--43.8, $>$43.8].
These bins have median values L$_{\textrm{\small \NII }}$/L$_{\textrm{\small FIR}}$ of $\sim 7.3\times 10^{-5}$, $3.1\times 10^{-5}$, and $3.0\times 10^{-5}$.

\section{Discussion}

Our \NII\ observations for a sample of 40 SMGs at $z=3-6$ represent the largest uniformly selected sample of high-redshift \NII\ detections to date.
In our 30 sources with both \NII\ and \CII\ observations, we are able to characterize the \ratioCN{} out to the high-redshifts probed by our SMG sample, and in over a decade range in (de-magnified) far-IR luminosity ($\sim12<\log_{10}($L$_{\textrm{\FIR }}/\mu)<13.5$).
This has allowed us to capture the true \ratioCN{} range for luminous, dusty galaxies, and better understand the outliers to the distribution.

All previous literature measurements of \ratioCN{} in distant, far-IR luminous galaxies are found to lie within the range of \ratioCN\ ratios we observe in the SPT SMGs.
The SPT SMGs exhibit among the highest and lowest ratios yet seen in high-redshift, FIR luminous galaxies.
However, our measurements also detect a population of high-redshift SMGs which have lower \ratioCN\ ratios than shown in previous literature.
We find 8 of our 30 SMGs have \ratioCN\ values (or are consistent within error) suggesting a hybrid regime between the model predictions of PDR/XDR emission and \HII\ regions.
These sources suggest neither \HII\ regions nor PDR/XDR regions dominate the \CII\ flux. 
Instead the \CII\ flux has significant contributions from both regions.
The \ratioCN\ in these sources cannot be explained as originating in only \HII\ regions -- both neutral and ionized gas must contribute to the total \CII\ 158 $\mu$m luminosity.
SMGs lying in this hybrid or even \HII\ dominated regimes could represent very enriched, high-metallicity gas in combination with low gas densities, and appropriate ionization parameters.
The C/N abundance ratio may vary as a function of metallicity (e.g. \protect\citealt{nagao12}), and could be an important contributor to the \ratioCN\ line ratio.
These SMGs may also have very high masses of ionized gas relative to the molecular gas fraction.
Detailed multi-line studies of these sources will be undertaken in follow-up contributions to better understand this unusual situation for such massive and dusty star-forming galaxies.

At the highest values of \ratioCN\ we detect 5 sources with larger ratios than the previous SMG record holder SPT2132--58 \protect\citep{bethermin16}, and comparable to the lower far-IR luminous sources of \protect\cite{umehata17} and \protect\cite{pavesi16}.
These sources likely represent extreme ISM environments, where total \CII\ emission is dominated by the contribution from PDR/XDR regions.
Since neutral nitrogen has a higher ionization potential than hydrogen, we expect that \NII\ 205 $\mu$m emission will only originate in ionized gas.
Therefore, our sources with extremely high \ratioCN\ may have a relatively low contribution to \CII\ from ionized gas where the \NII\ originates.

Comparing our SMG sample to local LIRGs, we observe slightly higher \ratioCN\ values, possibly owing to lower density gas and therefore a lower contribution of \CII\ emission from PDR/XDR regions in local (U)LIRGs.
However, the \textit{p}-values from both the T-test and KS-test (see Section~\ref{sec:results}) suggest we cannot say the SMGs and local LIRGs arise from different underlying distributions.

We follow the analysis of \protect\cite{nagao12} and include two metallicity grids for gas densities of $\log(n)=1.5$ and $=3$, each with ionization parameter $\log(U)=-3.5$ in Figure~\ref{fig:fig3}. According to the models of \protect\cite{nagao12}, higher \ratioCN\ should originate in lower metallicity environments. In our sample, we see that the majority of our sample of galaxies have line ratios that place them in the metallicity range of $0.6<Z/Z_{\odot}<3.0$ if we assume the $\log(n)=3$. Assuming $\log(n)=1.5$, we find metallicities in the range $0.2<Z/Z_{\odot}<1.0$. In both cases, the metallicity spans sub- to solar or super-solar ranges.
However, direct interpretation of \ratioCN\ in terms of metallicity is undermined by unconstrained gas density, elemental abundances, fractional abundance of ionized and neural gas, and ionization parameter, which can also affect this luminosity ratio (e.g.\ \protect\citealt{nagao12}, \protect\citealt{pavesi16}). In galaxies with a significant fraction of neutral ISM, \CII\ emission will more heavily out-weigh \NII\ than in galaxies with a significant ionized ISM component.

We investigate the relationship between L$_{\textrm{\small \NII }}$/L$_{\textrm{\small FIR}}$ versus T$_{\textrm{\small dust}}$.
We observe a deficit in L$_{\textrm{\small \NII }}$/L$_{\textrm{\small FIR}}$ towards higher T$_{\textrm{\small dust}}$ after binning according to dust temperature.
\protect\cite{gullberg15} similarly observed decreasing L$_{\textrm{\small \CII }}$/L$_{\textrm{\small FIR}}$ towards increasing dust temperature.
This result was first presented and explained in \protect\cite{malhotra97} who explained this ratio may change due to one of two reasons: (1) high far ultraviolet flux to gas density ratios may positively charge dust grains and therefore decrease heating efficiency, or (2) softer radiation fields can be less effective in heating gas and instead heat only the dust.
\protect\cite{gullberg15} acknowledged the Stefan-Boltzmann law may explain part of this dependence, as L$_{\textrm{\small FIR}}\propto$ T$_{\textrm{\small dust}}^4$.
To cancel this relationship, we plot $($L$_{\textrm{\small \NII }}$/L$_{\textrm{\small FIR}})\times ($T$_{\textrm{\small dust}}/40$ K$)^4$ versus T$_{\textrm{\small dust}}$ in Figure~\ref{fig:fig4}.
After removing this dependence, we perform a Kendall Tau test on the sample and calculate a \textit{p}-value of 0.49.
Similarly to the results presented in \protect\cite{gullberg15}, this result means we neither confirm the existence of a correlation between these variables, nor that \NII\ emission is largely independent of dust temperature for our SPT SMGs.
To investigate whether the trend appears in sources depending on the relative significance of \NII\ emission, we reduce the sample into two sub-samples according to their \ratioCN{}.
We define the \NII\ significant sources as those with \ratioCN{}$<10$, and the \NII\ insignificant sources with \ratioCN{}$>10$.
For these samples, we repeat the Kendall Tau test and calculate \textit{p}-values of 0.17 and 0.08, respectively.
This indicates our sources with significant ionized gas emission \ratioCN{}$<10$ do not exhibit a correlation between L$_{\textrm{\small \NII }}$/L$_{\textrm{\small FIR}}$ and T$_{\textrm{\small dust}}$, nor do sources with more significant PDR/XDR region emission.

\section{Conclusions}

We have presented the first uniformly selected sample of high-redshift \NII\ 205 $\mu$m observations and utilized previous observations of the \CII\ 158 $\mu$m line to probe the ISM.
We summarize our main conclusions here:

\begin{itemize}

\item{
We find that our SPT SMGs have a wide distribution of \ratioCN\/.
The median \ratioCN\ is \median\ with an interquartile range of \interquartile\/.
Using survival statistics to account for our lower limits did not significantly alter our results.
This resulted in a median of \survmedian\ and an interquartile range of \survinterquartile\/.
}

\item{
We measure a decrease in L$_{\textrm{\small \NII }}$/L$_{\textrm{\small FIR}}$ towards increasing L$_{\textrm{\small FIR}}$.
From the lowest luminosity bin (log$_{10}($L$_{\textrm{\small FIR}})<12.71$) to our highest luminosity bin (log$_{10}($L$_{\textrm{\small FIR}})>12.9$) we find medians of $\sim 1.6\times 10^{-4}$ decreasing to $\sim 1.2\times 10^{-4}$.
}

\item{
We cannot determine whether our measured \NII\ emission is independent of dust temperature, after cancelling the L$_{\textrm{\small FIR}}\propto$ T$_{\textrm{\small dust}}^4$ dependence and performing a Kendall Tau correlation test, including samples with high and low \ratioCN\ values.
}

\item{
Our range in observed \ratioCN\ can be explained through variations in gas density, ionization parameter, and metallicity.
We note that further observations of fine-structure lines such as \NII\ 122 $\mu$m and \OI\ 145 $\mu$m will help break the degeneracy of the \ratioCN\ on gas density and ionization parameter (\protect\citealt{nagao12}), and will help strengthen conclusions based on comparisons of the \CII\/-to-\NII\ ratio between local LIRGs and the high-redshift universe.
}

\end{itemize}

\input{table_full_flux_order}

\label{section:red}

\section*{Acknowledgements}

This paper makes use of the following ALMA data: ADS/JAO.ALMA\#2016.1.00133.T. ALMA is a partnership of ESO (representing its member states), NSF (USA) and NINS (Japan), together with NRC (Canada), MOST and ASIAA (Taiwan), and KASI (Republic of Korea), in cooperation with the Republic of Chile. The Joint ALMA Observatory is operated by ESO, AUI/NRAO and NAOJ.

The SPT is supported by the NSF through grant PLR-1248097, with partial support through PHY-1125897, the Kavli Foundation and the Gordon and Betty Moore Foundation grant GBMF 947.
D.P.M. J.D.V., K.C.L. and S.J. acknowledge support from the US NSF under grants AST-1715213 and AST-1716127.
S.J. and K.C.L acknowledge support from the US NSF NRAO under grants SOSPA5-001 and SOSPA4-007, respectively.
J.D.V. acknowledges support from an A. P. Sloan Foundation Fellowship.

The National Radio Astronomy Observatory is a facility of the National Science Foundation operated under cooperative agreement by Associated Universities, Inc.

D.J.M.C and S.C.C. acknowledge the support of the Natural Sciences and Engineering Research Council of Canada (NSERC).

\nocite{*} 
\bibliographystyle{mnras}
\bibliography{biblio}

\end{document}

%% file: table_full_flux_order.tex
\begin{table*}
\caption[dum]{Observed properties for our 41 SMGs ordered by $\mu$ S$_{\textrm{\small \NII }}$, the observed line flux without gravitational lensing correction. The \NII\ properties are determined from a single pixel extracted \NII\ spectra. The instrinic flux is determined by dividing the observed flux by the gravitational lensing factor $\mu$. Redshifts are provided from \protect\cite{strandet16}, while lensing models and $\mu$ values can be found in \protect\cite{spilker16}. For sources without lensing models, we assume our median magnification of $\mu = 6.3$ for strongly lensed SMGs \protect\citep{spilker16}. Sources with model magnifications close to or equal to one are likely unlensed systems, some suspected to be the cores of dense protoclusters (e.g.\ \protect\citealt{miller18}). Sources without a FWHM listed in the \CII\ column do not have \CII\ observations. Sources without a FWHM listed in the \NII\ column have line profiles too noisy to reliably fit with a gaussian function.}
\label{table:tab1}
\begin{tabular}{ccccccccc}\hline
\multicolumn{1}{c}{Source} & \multicolumn{1}{c}{$z$} & \multicolumn{1}{c}{$\mu$ S$_{\textrm{\small \NII }}{}^{a}$} & \multicolumn{1}{c}{SNR$_{\textrm{\small \NII }}{}^{b}$} &  \multicolumn{1}{c}{$\mu$ L$_{\textrm{\small \NII }}$} & \multicolumn{1}{c}{$\mu$ S$_{\textrm{\small \CII }}$} & \multicolumn{1}{c}{FWHM$_{\textrm{\small \NII }}$} & \multicolumn{1}{c}{FWHM$_{\textrm{\small \CII }}$} & \multicolumn{1}{c}{$\mu$}\\
\multicolumn{1}{c}{} & \multicolumn{1}{c}{} & \multicolumn{1}{c}{(Jy km~s$^{-1}$)} & \multicolumn{1}{c}{} & \multicolumn{1}{c}{($\times$10$^{8}$~$L_{\odot}$)} & \multicolumn{1}{c}{(Jy km~s$^{-1}$)} & \multicolumn{1}{c}{(km~s$^{-1}$)} & \multicolumn{1}{c}{(km~s$^{-1}$)} & \multicolumn{1}{c}{}\\
\hline
0529-54 & 3.3689 & 13.6 $\pm$ 1.7 & 9.3 & 40.4 $\pm$ 5.0 & 64.6 $\pm$ 7.7 & 415 $\pm$ 44 & 733 $\pm$ 81 & 13.2 $\pm$ 0.8\\
0103-45 & 3.0917 & 12.1 $\pm$ 2.3 & 11.7 & 31.2 $\pm$ 5.9 & 190.4 $\pm$ 23.8 & 429 $\pm$ 67 & 239 $\pm$ 39 & 5.1 $\pm$ 0.1\\
0155-62 & 4.349 & 11.6 $\pm$ 1.0 & 5.7 & 51.6 $\pm$ 4.5 & 26.3 $\pm$ 6.2 & 732 $\pm$ 104 & 760 $\pm$ 127 & 6.3 $\pm$ 1.0$^f$\\
2146-55 & 4.5672 & 9.3 $\pm$ 0.7 & 12.7 & 44.6 $\pm$ 3.4 & 11.3 $\pm$ 5.8 & 442 $\pm$ 55 & 277 $\pm$ 73 & 6.6 $\pm$ 0.4\\
0113-46 & 4.2328 & 9.1 $\pm$ 0.5 & 8.7 & 38.8 $\pm$ 2.1 & 49.7 $\pm$ 12.8 & 616 $\pm$ 65 & 578 $\pm$ 136 & 23.9 $\pm$ 0.5\\
0532-50 & 3.3988 & 9.0 $\pm$ 0.8 & 7.3 & 27.1 $\pm$ 2.4 & 91.7 $\pm$ 11.2 & 755 $\pm$ 123 & 719 $\pm$ 125 & 10.0 $\pm$ 0.6\\
2357-51 & 3.0703 & 8.7 $\pm$ 1.1 & 7.3 & 22.2 $\pm$ 2.8 & 9.1 $\pm$ 4.6 & 692 $\pm$ 129 & 743 $\pm$ 202 & 2.9 $\pm$ 0.1\\
2037-65 & 4.000 & 8.4 $\pm$ 0.7 & 8.1 & 32.8 $\pm$ 2.7 & --- & 503 $\pm$ 76 & --- & 6.3 $\pm$ 1.0$^f$\\
0109-47 & 3.6137 & 8.1 $\pm$ 1.8 & 9.1 & 26.9 $\pm$ 6.0 & --- & 545 $\pm$ 75 & --- & 10.2 $\pm$ 1.0\\
2147-50 & 3.7602 & 7.9 $\pm$ 0.6 & 9.1 & 28.0 $\pm$ 2.1 & 24.4 $\pm$ 7.5 & 570 $\pm$ 63 & 534 $\pm$ 104 & 6.6 $\pm$ 0.4\\
0020-51 & 4.1228 & 7.5 $\pm$ 0.6 & 9.4 & 30.7 $\pm$ 2.5 & --- & 459 $\pm$ 66 & --- & 4.2 $\pm$ 0.1\\
0027-50 & 3.4436 & 7.3 $\pm$ 0.9 & 8.2 & 22.5 $\pm$ 2.8 & --- & 288 $\pm$ 38 & --- & 5.1 $\pm$ 0.2\\
0418-47 & 4.2248 & 7.2 $\pm$ 0.6 & 12.5 & 30.6 $\pm$ 2.6 & 138.1 $\pm$ 10.4 & 366 $\pm$ 35 & 322 $\pm$ 37 & 32.7 $\pm$ 0.7\\
2103-60 & 4.4357 & 5.8 $\pm$ 0.6 & 8.7 & 26.6 $\pm$ 2.8 & 15.6 $\pm$ 10.4 & 670 $\pm$ 110 & 602 $\pm$ 204 & 27.8 $\pm$ 1.8\\
0459-59 & 4.7993 & 5.5 $\pm$ 0.8 & 8.2 & 28.5 $\pm$ 4.1 & --- & 464 $\pm$ 63 & --- & 4.2 $\pm$ 0.4\\
0551-50 & 3.164 & 5.2 $\pm$ 1.3 & 5.9 & 13.9 $\pm$ 3.5 & 216.1 $\pm$ 16.3 & 775 $\pm$ 273 & 734 $\pm$ 95 & 4.5 $\pm$ 0.5$^c$\\
2101-60 & 3.156 & 5.0 $\pm$ 1.8 & 6.0 & 13.3 $\pm$ 4.8 & 9.3 $\pm$ 13.8 & 682 $\pm$ 273 & 353 $\pm$ 189 & 6.3 $\pm$ 1.0$^f$\\
0136-63 & 4.299 & 4.8 $\pm$ 0.7 & 3.8 & 21.0 $\pm$ 3.1 & 33.3 $\pm$ 2.9 & 500 $\pm$ 128 & 526 $\pm$ 115 & 6.3 $\pm$ 1.0$^f$\\
0125-50$^d$ & 3.959 & 4.7 $\pm$ 0.9 & 4.2 & 18.1 $\pm$ 3.5 & --- & --- & --- & 14.1 $\pm$ 0.5\\
0345-47 & 4.2958 & 4.4 $\pm$ 0.7 & 4.9 & 19.2 $\pm$ 3.1 & 15.4 $\pm$ 4.3 & 350 $\pm$ 75 & 669 $\pm$ 177 & 8.0 $\pm$ 0.5\\
2048-55 & 4.089 & 4.4 $\pm$ 1.1 & 3.0 & 17.8 $\pm$ 4.4 & --- & 472 $\pm$ 156 & --- & 6.3 $\pm$ 0.7\\
0550-53$^d$ & 3.128 & 4.2 $\pm$ 1.4 & 4.5 & 11.0 $\pm$ 3.7 & 88.1 $\pm$ 8.6 & --- & 789 $\pm$ 165 & 6.3 $\pm$ 1.0$^f$\\
0459-58 & 4.856 & 4.1 $\pm$ 0.7 & 4.0 & 21.6 $\pm$ 3.7 & --- & 525 $\pm$ 143 & --- & 5.0 $\pm$ 0.6\\
2311-54 & 4.2795 & 3.6 $\pm$ 0.6 & 3.9 & 15.6 $\pm$ 2.6 & 45.3 $\pm$ 4.6 & 315 $\pm$ 129 & 352 $\pm$ 52 & 6.3 $\pm$ 1.0$^f$\\
0243-49$^d$ & 5.699 & 3.3 $\pm$ 0.9 & 3.9 & 22.1 $\pm$ 6.0 & 17.4 $\pm$ 2.7 & --- & 796 $\pm$ 202 & 6.7 $\pm$ 0.5\\
0348-62 & 5.656 & 3.0 $\pm$ 0.6 & 4.8 & 19.9 $\pm$ 4.0 & 19.6 $\pm$ 3.4 & 507 $\pm$ 213 & 506 $\pm$ 132 & 1.2 $\pm$ 0.01\\
2340-59 & 3.864 & 2.6 $\pm$ 0.5 & 4.5 & 9.6 $\pm$ 1.8 & 48.1 $\pm$ 8.5 & 579 $\pm$ 198 & 473 $\pm$ 220 & 3.4 $\pm$ 0.3\\
0516-59 & 3.4045 & 2.6 $\pm$ 0.8 & 3.8 & 7.9 $\pm$ 2.4 & --- & 156 $\pm$ 254 & --- & 6.3 $\pm$ 1.0$^f$\\
0245-63 & 5.626 & 2.4 $\pm$ 0.2 & 2.9 & 15.8 $\pm$ 1.3 & 26.9 $\pm$ 4.5 & 241 $\pm$ 83 & 383 $\pm$ 65 & 1.0 $\pm$ 0.01\\
2353-50 & 5.576 & 2.2 $\pm$ 0.3 & 2.6 & 14.3 $\pm$ 1.9 & 20.7 $\pm$ 6.4 & 613 $\pm$ 405 & 429 $\pm$ 247 & 6.3 $\pm$ 1.0$^f$\\
0300-46$^d$ & 3.5954 & 2.1 $\pm$ 1.2 & 4.2 & 6.9 $\pm$ 4.0 & 15.7 $\pm$ 5.1 & --- & 414 $\pm$ 237 & 5.7 $\pm$ 0.4\\
2319-55 & 5.2929 & 2.0 $\pm$ 0.8 & 3.9 & 12.0 $\pm$ 4.8 & 39.2 $\pm$ 4.7 & 339 $\pm$ 245 & 176 $\pm$ 28 & 6.9 $\pm$ 0.6\\
0319-47 & 4.51 & 2.0 $\pm$ 0.5 & 3.5 & 9.4 $\pm$ 2.4 & 11.4 $\pm$ 10.5 & 259 $\pm$ 42 & 562 $\pm$ 182 & 2.9 $\pm$ 0.3\\
2307-50 & 3.105 & 1.9 $\pm$ 0.6 & 3.2 & 4.9 $\pm$ 1.6 & --- & $\sim 98$ & --- & 6.3 $\pm$ 1.0$^f$\\
2132-58$^e$ & 4.7677 & 1.7 $\pm$ 0.2 & 11.5 & 8.9 $\pm$ 0.8 & 35.9 $\pm$ 6.9 & 245 $\pm$ 16 & 212 $\pm$ 43 & 5.7 $\pm$ 0.5\\
0441-46 & 4.4771 & 1.5 $\pm$ 0.5 & 2.9 & 7.0 $\pm$ 2.3 & 26.3 $\pm$ 5.8 & 393 $\pm$ 106 & 546 $\pm$ 123 & 12.7 $\pm$ 1.0\\
2349-56 & 4.304 & 1.1 $\pm$ 0.5 & 2.9 & 4.8 $\pm$ 2.2 & --- & --- & --- & 1.0\\
0202-61 & 5.018 & 0.036$^g$ & 4.5 & --- & 19.3 $\pm$ 4.1 & $\sim 60$ & 771 $\pm$ 325&  9.1 $\pm$ 0.07\\
0346-52 & 5.6559 & $>$0.9 & 2.9 & $>$6.0 & 64.1 $\pm$ 8.2 & $\sim 64$ & 486 $\pm$ 85 & 5.6 $\pm$ 0.1\\
2052-56 & 4.257 & 0.1 $\pm$ 0.4 & 2.3 & 0.4 $\pm$ 1.7 & 14.6 $\pm$ 1.8 & $\sim 91$ & 382 $\pm$ 122 & 1.0\\
2351-57 & 5.811 & $>$-0.8 & 2.7 & $>$-5.5 & 5.4 $\pm$ 2.7 & 383 $\pm$ 409 & 539 $\pm$ 82 & 6.3 $\pm$ 1.0$^f$\\

\hline
\end{tabular}

\noindent$^a$ Errors quoted are RMS from the 1D spectra.\\
\noindent$^b$ SNR determined from the source peak in \textit{uv}-plane continuum subtracted channel maps.\\
\noindent$^c$ Lens model by K. Sharon, private communication.\\
\noindent$^d$ The \NII\ spectra for these sources does not have sufficient baseline to completely cover the CO line profile. The fluxes quoted for \NII\ and \CII\ are truncated to the \NII\ spectral coverage.\\
\noindent$^e$ from \protect\cite{bethermin16}.\\
\noindent$^f$ $\mu=6.3$ is assumed. No lensing model.\\
\noindent$^g$ Positive flux measurement from channel map (Jy/beam). Our procedure for measuring total flux results in a negative value for this source. \\

\end{table*}